\documentclass[review]{elsarticle}
\usepackage{lineno,hyperref}
\modulolinenumbers[5]
\usepackage{color}
\usepackage{xspace}
\makeatletter
\DeclareRobustCommand{\LaTeX}{L\kern-.26em%
        {\sbox\z@ T%
         \vbox to\ht\z@{\hbox{\check@mathfonts
           \fontsize\sf@size\z@
           \math@fontsfalse\selectfont
          A\,}%
         \vss}%
        }%
     \kern-.15em%
    \TeX}
\makeatother
\def\edth{\;\raise1.0pt\hbox{$'$}\hskip-6pt\partial\;}
\def\baredth{\;\overline{\raise1.0pt\hbox{$'$}\hskip-6pt
\partial}\;}
\def\gsim{~\rlap{$>$}{\lower 1.0ex\hbox{$\sim$}}}

\newcommand{\be}{\begin{equation}}
\newcommand{\ba}{\begin{eqnarray}}
\newcommand{\ee}{\end{equation}}
\newcommand{\ea}{\end{eqnarray}}

\newcommand{\fr}{\frac}
\newcommand{\tF}{\tilde{F}}
\begin{document}
\title{Axion Dark Matter Induced Cosmic Microwave Background $B$ modes}
\author[tku]{Guo-Chin Liu}
\author[iop,iaa]{Kin-Wang Ng}
\address[tku]{Department of Physics, Tamkang University, Tamsui, New Taipei City 25137, Taiwan}
\address[iop]{Institute of Physics, Academia Sinica, Taipei 11529, Taiwan}
\address[iaa]{Institute of Astronomy and Astrophysics, Academia Sinica, Taipei 11529, Taiwan}
\begin{abstract}
It was known that isocurvature perturbation of a nearly massless cosmological axion field can lead to rotation of $E$-mode polarization into $B$-mode polarization in the cosmic microwave background (CMB) by the presence of a parity violating coupling of the field to the topological density of electromagnetism, resulting in a phenomenon known as anisotropic cosmic birefringence.
In this {\em Letter}, we propose a new source of anisotropic cosmic birefringence induced by dark matter adiabatic density perturbation.
If dark matter is ultralight axions that carry a coupling to photon, its adiabatic density fluctuations will induce anisotropic cosmic birefringence with a blue-tilted rotation power spectrum, thus generating CMB $B$-mode polarization on sub-degree angular scales. Using current POLARBEAR and SPTPol $B$-mode polarization data, we derive a constraint on the axion-photon coupling strength ($\beta$) and the axion mass ($m$),  $\beta^2 (10^{-22}{\rm eV}/m)^2 <  8\times 10^{15}$. It is shown that the birefringence $B$ modes can dominate over CMB lensing $B$ modes at high $l$, manifesting as an excess power for $l>1500$ in future CMB lensing $B$-mode searches. In addition, we derive the lensing-rotation cross correlation that can be a potential test to the present model. 

\end{abstract}
\maketitle

It is compelling that the present Universe is filled with dark matter (DM)~\cite{planck}.
Although the nature of DM remains elusive,
its gravitational pull is essential to the formation of large-scale structures.
It has been successfully modeled as massive weakly interacting particles or cold dark matter (CDM).
However, there exist serious discrepancies between observations and numerical simulations of CDM halos,
which predict too much power on small scales, manifested as cuspy CDM cores in dwarf galaxies,
galaxies like the Milky Way, and central regions of galaxy clusters
as well as a large excess of CDM subhalos or dwarf galaxies~\cite{primack}.
These discrepancies, if true, would suggest a suppressed matter power spectrum at small scales.
The power suppression can be achieved in many DM models, such as warm, self-interacting, decaying, and ultralight axion (ULA) DM. It is expected that future observations of small-scale structures may distinguish between these DM models (see, for examples, Refs.~\cite{viel,elbert,cheng,hlozek}). On the other side, the feedback of baryonic processes~\cite{pontzen14} and the effect of tidal stripping~\cite{fattahi16} can significantly alter the DM distribution, thus alleviating the small-scale problem and avoiding exotic particle physics (for a review, see Ref.~\cite{weinberg14}).

Perhaps the most motivated are the axion DM models. The best well-known is the QCD axion that was originally invented to solve the strong CP problem; as a bonus, it is a viable DM candidate~\cite{pdg}. Recently, it was proposed that string theory
suggests the presence of a plenitude of axions, possibly populating each decade of mass down to the
Hubble scale~\cite{axiverse}. As long as the condition $m > 3H$, where $m$ is the axion mass and $H$ is the Hubble parameter, is satisfied, the axion begins to coherently oscillate with an amplitude set by its initial vacuum expectation value (vev). This constitutes a homogeneous condensate with its energy density redshifting as $a^{-3}$ (where $a$ is the cosmic scale factor). If $m > 10^{-27}{\rm eV}$, the axion condensate behaves just like CDM after matter-radiation equality. Moreover, for ULAs with masses $m < 10^{-20}{\rm eV}$, the de Broglie wave can suppress small-scale power on astronomically observable length scales~\cite{hu,hlozek,huilam}. In numerical calculations of the formation of large-scale structures using the axion field with $m \sim 10^{-22}{\rm eV}$, it was shown that the ULA DM model may offer a viable solution to the small-scale problems~\cite{hu,marsh}. However, more recent constraints on $m$ from the core radii of dwarf spheroidal galaxies~\cite{chen16,morales16} have suggested a smaller ULA mass which is in tension with the mass found in large-scale-structure simulations~\cite{morales16}.

In this {\em Letter}, we assume DM to be an ULA field, $\Psi\equiv M\psi$, that couples to the electromagnetic field strength via
$(-\beta/4)\psi F_{\mu\nu} \tF^{\mu \nu}$, where $\beta$ is a coupling constant and $M$ is the reduced Planck mass.
We will not specify the particle model for the axion except assuming the axion mass to be around $10^{-22}{\rm eV}$.
For such an ULA, the most stringent upper bound on $\beta$ comes from the absence of a $\gamma$-ray burst in coincidence with
Supernova 1987A neutrinos, which would have been converted in the galactic magnetic field
from a burst of axion-like particles due to the Primakoff production in the supernova core:
$\beta < 2.4\times 10^7$~\cite{sn1987}.
The effect of this coupling to cosmic microwave background (CMB) polarization has been previously studied in many contexts such as new high-energy physics~\cite{pospelov}, a massless pseudo-Nambu-Goldstone spectator field~\cite{Caldwell}, and scalar quantum fluctuations of the vacuum-like cosmological constant~\cite{LLN2,Zhao}. It is well known that the above $\psi$-photon interaction leads to cosmic birefringence~\cite{carroll} that induces a rotation of the polarization plane of the CMB, thus converting $E$-mode into $B$-mode polarization without affecting the temperature anisotropy~\cite{lue,LLN}.
For the ULA being the DM component, we consider the contribution of $\psi$ perturbation to the cosmic birefringence fluctuations.

We adopt a flat geometry, $ds^2=a^2(\eta) (d\eta^2- d \vec{x}^2)$,
where $a(\eta)$ is the cosmic scale factor and $\eta$ is the conformal time
defined by $dt=a(\eta)d\eta$.
The $\psi F \tF$ term causes a rotational speed of the polarization plane of a photon
propagating in the direction $\hat{n}$~\cite{carroll},
\be
\omega(\vec{x},\eta)=-{\beta\over2}\left(\fr{\partial \psi}{\partial\eta}+
\vec{\nabla} \psi \cdot \hat{n}\right).
\label{rot}
\ee
Thomson scatterings of anisotropic CMB photons by free electrons
give rise to linear polarization, which can be described
by the Stokes parameters $Q(\vec{x},\eta)$ and $U(\vec{x},\eta)$.
The time evolution of the linear polarization is governed by the collisional Boltzmann equation,
which would be then modified due to the rotational speed of the polarization plane~(\ref{rot}) by including
a temporal rate of change of the Stokes parameters:
\be
\dot Q\pm i \dot U = \mp i2\omega \left( Q\pm i U \right),
\label{QUeq}
\ee
where the dot denotes $d/d\eta$.
This can be accounted as a convolution of the Fourier modes of the Stokes parameters
with the spectral rotation that can be easily incorporated into the Boltzmann code.

Now let us consider the time evolution of $\psi$.
We split $\psi$ into the mean field or the vev and the perturbation:
$\psi(\vec{x},\eta) =\bar{\psi}(\eta) + {\cal\psi} (\vec{x},\eta)$.
When the ULA begins to oscillate at $a=a_{\rm osc}$, the energy density of the ULA condensate is given by
\begin{equation}
\rho=m^2 M^2\bar{\psi}^2=m^2 M^2\bar{\psi}_i^2 \left(\frac{a_{\rm osc}}{a}\right)^3,
\end{equation}
where $\bar{\psi}_i$ is the initial vev. Hence, the perturbation is
\begin{equation}
{\cal\psi}={1\over2} \bar{\psi} \frac{\delta\rho}{\rho}
=\frac{\sqrt{3}}{2}\Omega_{\rm DM}^{1\over2} \frac{H_0}{m} (1+z)^{3\over2} \frac{\delta\rho}{\rho},
\end{equation}
where $\Omega_{\rm DM}$ and $H_0$ each take the present values when $a=a_0=1$.
Here $\delta\equiv {\delta\rho}/{\rho}$ is assumed to be the adiabatic DM density perturbation.
In terms of their perturbation power spectra, we have
\begin{equation}
\Delta_{\cal\psi}^2(k,\eta)
={3\over 4}\Omega_{\rm DM} \left(\frac{H_0}{m}\right)^2 (1+z)^3 \Delta_\delta^2(k,\eta) ,
\end{equation}
The DM power spectrum is related to that of the gravitational potential $\Phi$ by the Poisson equation:
\begin{equation}
\Delta_\Phi^2(k,\eta)={9\over4}\left(\frac{H_0}{k}\right)^4 \Omega_{\rm DM}^2 (1+z)^2 \Delta_\delta^2(k,\eta),
\end{equation}
where we have neglected baryons in the matter density.

It is useful to observe the rotation angle $\alpha({\hat n})=\sum_{lm} \alpha_l^m Y_l^m ({\hat n})$~\cite{Li,Caldwell,LLN2}.
Using Eq.~(\ref{rot}), it can be estimated as
\be
\alpha_l^m= i^l{\beta\over\sqrt{2\pi}}\int d^3{\vec k}\, Y_{l}^{m*}(\hat k)\,
{\cal\psi}({\vec k},\eta_s) j_l[k(\eta_0-\eta_s)]\, ,
\label{alphalm}
\ee
where $\eta_0$ is the present time and $\eta_s$ denotes the epoch when the primary CMB polarization is generated
at the last scattering surface or the reionization surface. Hence the rotation power spectrum is given by
\be
C_l^{\alpha\alpha}=\left<|\alpha_l^m|^2\right>= \pi\beta^2
\int \frac{dk}{k} \Delta_{\cal\psi}^2(k,\eta_s) \, j_l[k(\eta_0-\eta_s)]^2\,,
\label{clalpha}
\ee

On the other hand, CMB photons traveling in the gravitational potential of the large scale structure are deflected by an angle given by the angular gradient of the projected potential~\cite{kaiser}, $\nabla\phi({\hat n})$, with
\begin{equation}
\phi({\hat n})=-2\int dD \frac{D_s-D}{D_s D} \Phi(D{\hat n},D),
\end{equation}
where $D$ is the comoving angular diameter distance,
\begin{equation}
D(z)=\int_0^z \frac{H_0}{H(z')} dz'.
\end{equation}
Expanding the lensing potential $\phi({\hat n})=\sum_{lm} \phi_l^m Y_l^m ({\hat n})$, we find
\begin{equation}
\phi_l^m=-\frac{4 i^l}{\sqrt{2\pi}} \int dD \frac{D_s-D}{D_s D}\int d^3{\vec k}\, Y_{l}^{m*}(\hat k)\,
\Phi({\vec k},D) j_l(kD)\,.
\label{philm}
\end{equation}
Hence, the lensing power spectrum is given by
\begin{eqnarray}
C_l^{\phi\phi}&=&\left<|\phi_l^m|^2\right>=4\pi\int \frac{dk}{k} \times
\nonumber \\
&&\left[-2\int dD \frac{D_s-D}{D_s D} \Delta_\Phi(k,D)\, j_l(kD/H_0)\right]^2\,.
\label{clphi}
\end{eqnarray}
In the present consideration both the gravitational lensing and the cosmic birefringence originate from the same matter density perturbation, so there is a lensing-rotation cross correlation that can be measured by making use of Eqs.~(\ref{alphalm}) and~(\ref{philm}) as
\begin{eqnarray}
C_l^{\phi\alpha}&=&\left<\phi_l^{m*}\alpha_l^m\right>=\left<\phi_l^{m}\alpha_l^{m*}\right> \nonumber \\
&=& 2\pi\beta \int \frac{dk}{k} \Delta_{\cal\psi}(k,\eta_s) \, j_l[k(\eta_0-\eta_s)]
\times \nonumber \\ &&
\left[-2\int dD \frac{D_s-D}{D_s D} \Delta_\Phi(k,D)\, j_l(kD/H_0)\right]\,.
\label{clphialpha}
\end{eqnarray}
This provides a new potential cross correlation between gravitational lensing and cosmic rotation, although the correlation should be rather small. It is because CMB polarization is generated in the reionization epoch or at the last scattering surface while lensing of the CMB takes place at relatively low redshifts. It would be very interesting to consider the lensing-rotation cross correlation for lensed polarized astrophysical sources.

The lensing and the rotation of the CMB polarization lead to a remapping of the primary polarization,
\begin{equation}
(\tilde{Q}\pm i\tilde{U}) (\hat n)= (Q\pm iU) (\hat n +\nabla\phi)\, e^{\mp i2\alpha(\hat n)}.
\end{equation}
Then, using the lensing and the rotation power spectra, respectively given in Eq.~(\ref{clphi}) and Eq.~(\ref{clalpha}), in the limit of weak lensing and small rotation angle, we can approximate the total induced $B$-mode polarization power spectrum by
\begin{eqnarray}
C_l^{BB}&=&\frac{1}{32\pi}\sum_{l_1,l_2} (2l_1+1)(2l_2+1) C_{l_1}^{EE}(\eta_s) \times \nonumber \\
&&\left( \begin{array}{ccc} l &l_1 &l_2\\  2 &-2 & 0\end{array} \right)^2 
\left\{ [1-(-1)^{l+l_1+l_2}]  [l_1(l_1+1) \right. \nonumber \\
&&\left. +l_2(l_2+1)-l(l+1)]^2 C_{l_2}^{\phi\phi} + 32 C_{l_2}^{\alpha\alpha} \right\} ,
\label{cbl}
\end{eqnarray}
where we have assumed primary $E$ modes only and the matrix is the Wigner 3-$j$ symbol. The first term is the well-known weak lensing $B$-mode~\cite{hu2} and the second is the birefringence $B$-mode~\cite{Li,Caldwell,LLN2}. Since the effects of the lensing and the rotation on the CMB polarization are orthogonal, the lensing-rotation cross correlation~(\ref{clphialpha}) is absent in Eq.~(\ref{cbl}) to all orders . Also, note that both $C_l^{TB}$ and $C_l^{EB}$ power spectra vanish due to the fact that the ensemble averages $\left<{\cal\psi}\right>=\left<\phi\right>=0$. In practice, $C_l^{TB}$ and $C_l^{EB}$ are non-zero and can be used to estimate the rotation power spectrum in a local universe~\cite{kamionkowski09}.

In the following, we will adopt {\em Planck} 6-parameter LCDM model~\cite{planck} as the base cosmology to compute the $B$-mode polarization power spectrum~(\ref{cbl}).
For the lensing $B$-mode, we introduce a parameter $A_L$, which scales the $C_l^{\phi\phi}$ power spectrum at each point in parameter space,
and which is used to lens the CMB spectra. The base value of the lensing parameter is $A_L=1$. For the birefringence $B$-mode, we define the cosmic birefringence parameter
$A_{\rm CB}\equiv\beta^2 (10^{-22}{\rm eV}/m)^2$. Recent CMB $B$-mode data has revealed a significant level of galactic polarized dust emission on large angular scales~\cite{planckdust,b+p}. Here we follow the model motivated by the {\em Planck} results for the dust contribution at $150 {\rm GHz}$ with a $B$-mode power spectrum~\cite{planckdust}, 
\begin{equation}
\frac{l(l+1)}{2\pi} C^{BB}_{l\,{\rm dust}}=0.0118\, \left(l\over 80\right)^{-0.42} {\rm\mu K^2}.
\label{ps_dust}
\end{equation}

Now we compare our predicted power spectra with recent sub-degree-scale $B$-mode measurements made by POLARBEAR~\cite{polarbearB} as well as SPTpol~\cite{sptB}. We do not use ACTpol $B$-mode data~\cite{actB} because the data has relatively large uncertainties. The measurements of $B$-mode polarization at degree angular scales made by BICEP2/Keck Array~\cite{bicepB} are less sensitive to the lensing and cosmic birefringence $B$ modes, so they are only for the use of determining the dust $B$-mode power spectrum~(\ref{ps_dust}). Hence, we subtract the dust polarization foreground from the measured polarization power assuming the dust power spectrum~(\ref{ps_dust}), and then minimize the chi-square values over the 2-dimensional parameter space $(A_{\rm CB}, A_L)$:
\begin{eqnarray}
&&\chi^2(A_{\rm CB}, A_L) \nonumber \\
&=&\sum_{bb^{\prime}}{\left(C_{b}^{\rm Obs}-C_{b}^{\rm dust}-A_{\rm CB}C^{\rm CB}_b-A_L C_b^{\rm lens}\right)M^{-1}_{bb^\prime}}
\nonumber \\
& & \left(C_{b^\prime}^{\rm Obs}-C_{b^\prime}^{\rm dust}-A_{\rm CB}C^{\rm CB}_{b^\prime}-A_L C^{\rm lens}_{b\prime}\right),
\end{eqnarray}
where $M_{bb^\prime}$ is the covariance matrix. $C_{b}^{\rm Obs}$, $C_{b}^{\rm dust}$, $C^{\rm lens}_b$, and $C_{b}^{\rm CB}$ are the band power in band $b$ for the measured, the dust, the expected lensed, and the cosmic birefringence induced $B$-mode polarization, respectively. Each band power is constructed from the respective power spectrum convolved with the window function as $C_b=\sum_l w^b_l C_l$.

\begin{figure}[h]
\makebox{\includegraphics[width=\textwidth]{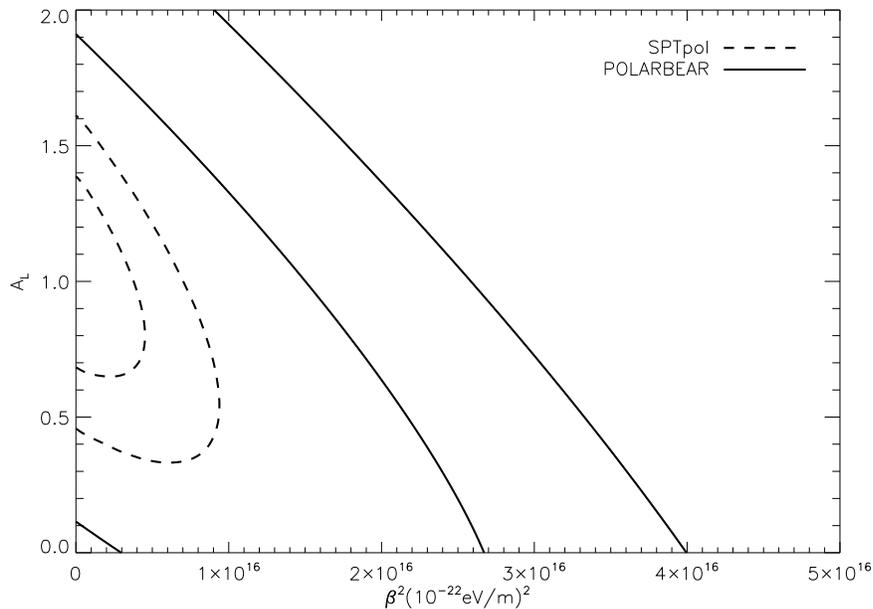}}
\caption{Likelihood plot of the parameters $A_L$ and $A_{\rm CB}$, showing $1$-sigma and $2$-sigma contours. Solid and dashed contours use POLARBEAR data~\cite{polarbearB} and SPTpol data~\cite{sptB}, respectively.}
\label{fig:contour}
\end{figure}

\begin{figure}[t]
\makebox{\includegraphics[width=\textwidth]{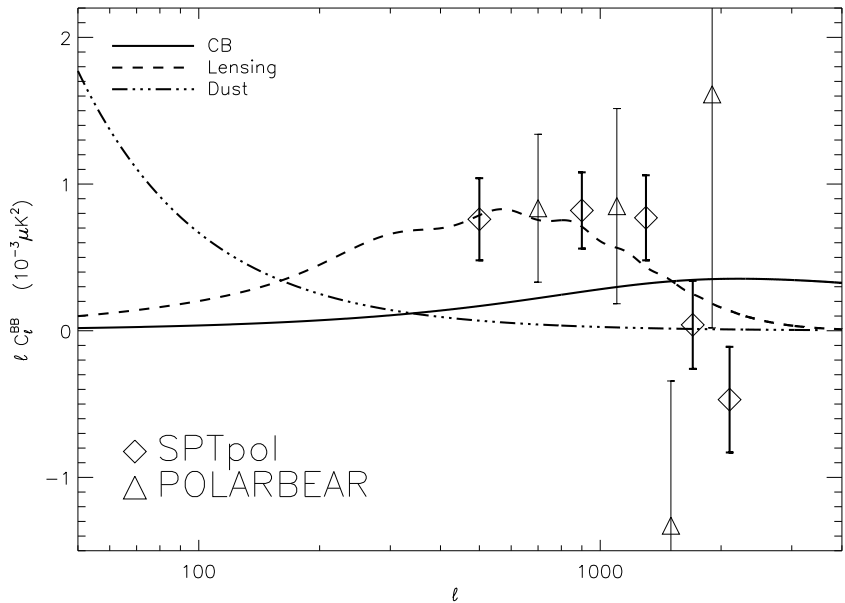}}
\caption{Cosmic birefringence induced $B$-mode power spectrum with $A_{\rm CB}=8\times 10^{15}$ (solid).
Also shown are the power spectra of lensing induced
$B$ modes with $A_L=1.07$ (dashed) and dust $B$ modes (dot-dashed). Overlaid are POLARBEAR data~\cite{polarbearB} (triangles) and SPTpol data~\cite{sptB} (diamonds).}
\label{fig:power}
\end{figure}

\begin{figure}[htbp]
\makebox{\includegraphics[width=\textwidth]{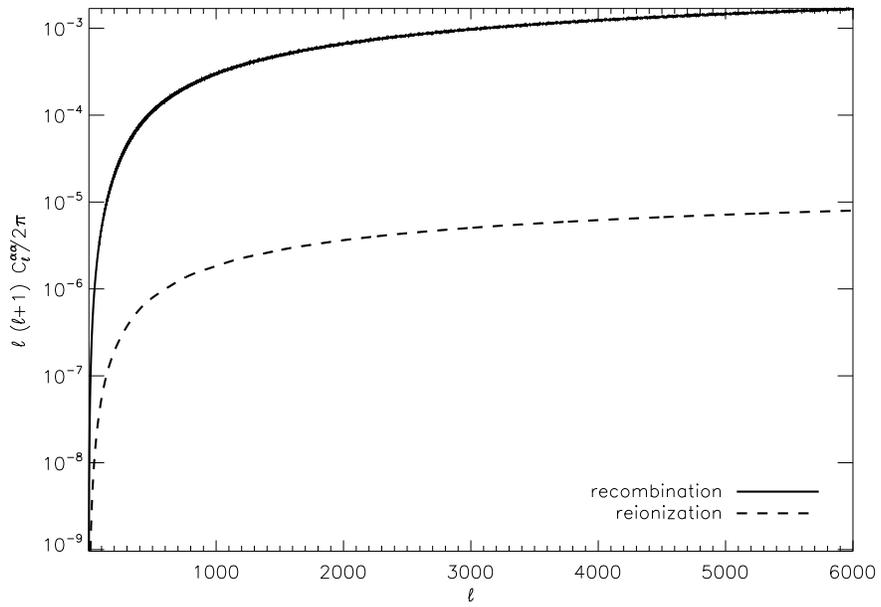}}
\caption{Rotation power spectra at the recombination and the reionization with 
$A_{\rm CB}=8\times 10^{15}$.}
\label{fig:alpha}
\end{figure}

In Fig.~\ref{fig:contour} we plot the likelihood contours of $A_L$ and $A_{\rm CB}$ inferred from the POLARBEAR data and the SPTpol data respectively. The two data sets give us consistent likelihood plots that have close center values and similar contour orientations. From the POLARBEAR data, we obtain the maximum likelihood value of $A_L=1.01$ and marginalizing $A_L$ gives the upper bound, $A_{\rm CB}<2.63\times 10^{16}$ (at $95\%$ c.l.). The SPTpol data gives the maximum likelihood value of $A_L=1.07$ and sets a tighter constraint with $A_{\rm CB}<8\times 10^{15}$ (at $95\%$ c.l.).  If an excess $B$-mode signal is detected at a level of $A_{\rm CB}=8\times 10^{15}$, then the astrophysical supernova bound, $\beta < 2.4\times 10^7$, would imply that $m<2.7\times 10^{-23}{\rm eV}$. Figure~\ref{fig:power} shows the lensing and birefringence $B$-mode power spectra with $A_L=1.07$ and $A_{\rm CB}=8\times 10^{15}$ respectively. The birefringence $B$ modes dominate the polarization power for $l>1400$; therefore, measurements of $B$-mode polarization at sub-degree scales are critical for probing cosmic birefringence induced by the ULA dark matter. Figure~\ref{fig:alpha} shows the blue-tilted rotation power spectra for the recombination and the reionization with $A_{\rm CB}=8\times 10^{15}$, which we have used to produce the birefringence $B$-mode power spectrum in Fig.~\ref{fig:power}.

Recently, constraints on the rotation power spectrum have been derived from WMAP 7-year data using $\langle TBTB \rangle$ four-point correlations~\cite{Gluscevic}, 
WMAP 9-year data using two-point correlation function~\cite{LXLLZ}, 
and POLARBEAR data using $\langle EBEB \rangle$ four-point correlations for a null test~\cite{polarbearCB}.
The POLARBEAR upper limit on the amplitude of a scale-invariant rotation power spectrum, $l(l+1) C_l^{\alpha\alpha}/(2\pi)= 10^{-4} A$, is given by $A<3.1$ 
(at $95\%$ c.l.)~\cite{polarbearCB}.
We have done the same likelihood analysis as above using a scale-invariant rotation power spectrum. From POLARBEAR $\langle BB \rangle$ power spectrum, we obtain $A<3.2$ when assuming no lensing ($A_L=0$), which is consistent with the POLARBEAR limit. Marginalizing $A_L$ gives a tighter upper limit, $A<2.07$, which is in accordance with the results in Ref.~\cite{polarbearCB}. From SPTpol $\langle BB \rangle$ power spectrum, we obtain the maximum likelihood value of $A_L=1.02$ and marginalizing $A_L$ gives $A<1.36$ (at $95\%$ c.l.), which is the tightest constraint on the scale-invariant rotation power spectrum. We stress that this constraint does not apply to the ULA induced rotation power spectrum because of its blue-tilted spectrum shape.

In conclusion, we have proposed a new source of CMB $B$-mode polarization induced by birefringence fluctuations of ultralight axion dark matter.
The power spectrum of this birefringence $B$-mode polarization peaks at sub-degree angular scales and may be at a level detectable in on-going CMB lensing $B$-mode searches such as ACTpol, POLARBEAR, and SPTpol experiments. Interestingly, it may dominate over the lensing $B$-mode power spectrum at higher-$l$ range. Thus, it would be very important to make precise measurements of $B$-mode polarization at sub-degree scales to disentangle the two $B$-mode signals. 
The current experimental sensitivity in measuring $l C_l^{BB}$ is of order $10^{-3} {\rm\mu K^2}$, which is at the same level of the $B$-mode signals~(see Fig.~\ref{fig:power}). 
In future CMB-S4 polarization experiments, the sensitivity will be tremendously improved to $\sim 10^{-6} {\rm\mu K^2}$ for $l<5000$~\cite{cmbs4}, so consistency of sub-degree $B$ modes with the lensing of $E$ modes will test the present ULA model in a well-defined way. Both $B$ modes originate from the same dark matter power spectrum, though converted differently from the CMB $E$-mode polarization, so they will display cut-off power spectra at the ULA Jeans scale ($l\sim 10^4$). In principle, one can use similar de-lensing methods~\cite{delensing} or lensing as well as rotation contributions to CMB bi-spectra~\cite{hu2}, assisted with the cross-correlation~(\ref{clphialpha}), to perform de-lensing and de-rotation simultaneously to separate them. More investigations along this line should be in order. 

We have also derived the cross correlation in Eq.~(\ref{clphialpha}) between the lensing and the rotation of CMB in the ULA model. The cross correlation can be equally applied to any astrophysical polarized source to provide an independent test of the ULA model. Furthermore, we have considered adiabatic DM density perturbation in this work. Including isothermal density perturbation would be an interesting extension.

This work was supported in part by the Ministry of Science and Technology, Taiwan, ROC under the Grants No. MOST104-2112-M-001-039-MY3 (K.W.N.) and No. MOST105-2112-M-032 -002 (G.C.L.).

\end{document}